\documentclass[pre,nofootinbib,amsmath,amssymb,preprint,showpacs]{revtex4-1}
\usepackage[T1]{fontenc} \usepackage[latin1]{inputenc}
\usepackage{amsmath} \usepackage{graphicx} \usepackage{amssymb}

\begin{document}

\title{Simulations of Two-Dimensional Unbiased Polymer Translocation
Using the Bond Fluctuation Model}

\author{Debabrata Panja} \affiliation{Institute for Theoretical
Physics, Universiteit van Amsterdam, Valckenierstraat 65, 1018 XE
Amsterdam, The Netherlands} \author{Gerard
T. Barkema$^{\dagger\ddagger}$} \affiliation{$^{\dagger} $Institute
for Theoretical Physics, Universiteit Utrecht, Leuvenlaan 4, 3584 CE
Utrecht, The Netherlands\\ $^{\ddagger}$Instituut-Lorentz,
Universiteit Leiden, Niels Bohrweg 2, 2333 CA Leiden, The Netherlands}

\begin{abstract} 
  We use the Bond Fluctuation Model (BFM) to study the pore-blockade
  times of a translocating polymer of length $N$ in two dimensions, in
  the absence of external forces on the polymer (i.e., unbiased
  translocation) and hydrodynamic interactions (i.e., the polymer is a
  Rouse polymer), through a narrow pore. Earlier studies using the BFM
  concluded that the pore-blockade time scales with polymer length as
  $\tau_d \sim N^\beta$, with $\beta=1+2\nu$, whereas some recent
  studies with different polymer models produce results consistent
  with $\beta=2+\nu$, originally predicted by us. Here $\nu$ is the
  Flory exponent of the polymer; $\nu=0.75$ in 2D. In this paper we
  show that for the BFM if the simulations are extended to longer
  polymers, the purported scaling $\tau_d \sim N^{1+2\nu}$ ceases to
  hold. We characterize the finite-size effects, and study the
  mobility of individual monomers in the BFM. In particular, we find
  that in the BFM, in the vicinity of the pore the individual
  monomeric mobilities are heavily suppressed in the direction
  perpendicular to the membrane. After a modification of the BFM which
  counters this suppression (but possibly introduces other artifacts
  in the dynamics), the apparent exponent $\beta$ increases
  significantly. Our conclusion is that BFM simulations do not rule
  out our theoretical prediction for unbiased translocation, namely
  $\beta=2+\nu$.
\end{abstract}


\maketitle

\section{Introduction\label{sec1}}

For polymer translocation through narrow pores in membranes, the
scaling behavior of pore-blockade times with the length of linear
polymers has been a topic of intense research in recent times. Such
interest in polymer translocation has been fueled by its obvious
biological context; i.e., molecular transport through cell membranes,
which is an essential mechanism in living organisms. Often, the
molecules are too long, and the pores in the membranes too narrow, to
allow the molecules to pass through as a single collapsed unit. In
such circumstances, the molecules have to deform themselves in order
to squeeze --- i.e., translocate --- through the pores. Parallely, the
urge to understand the dynamics of translocation also stems from the
fact that new developments in design and fabrication of
nanometer-sized pores and etching methods, in recent times, have put
translocation at the forefront of single-molecule experiments, with
the hope that translocation may lead to cheaper and faster technology
for the analysis of biomolecules.

Although significant progress has been made in the last few years in
the field of both theory and simulations of polymer translocation,
consensus among different research groups on the scaling behavior of
the characteristic time $\tau_d$ that the polymer spends in the pore,
with the length $N$ of a linear polymer, characterized by an exponent
$\beta \equiv \partial (\log(\tau_d))/\partial (\log(N))$, has
generally remained elusive. Of the three main translocation situations
studied theoretically or by computer simulations, namely (i) unbiased
translocation, wherein the polymer translocates purely due to thermal
fluctuations, (ii) field-driven translocation, wherein translocation
is driven by a potential difference across the pore, and (iii) pulled
translocation, wherein translocation is facilitated by a pulling force
at the head of the polymer, unbiased translocation is by far the most
fiercely debated topic.

From a statistical physics perspective, the translocation problem can
be seen as a kind of a tunneling process over an entropic barrier.
This entropic barrier arises because the number of states available to
the polymer is significantly decreased by the presence of the
membrane. For a polymer of length $N$, the number of states in the
bulk scales as $Z_b(N)\approx A~\mu^N ~N^{\gamma-1}$ in which $\gamma$
is a universal exponent --- $\gamma=49/32$ and $\gamma\approx 1.16$ in
two and three dimensions, respectively --- while $A$ and $\mu$ are not
universal. The corresponding number of states for the same polymer,
but whose end is tethered to the membrane, is approximated by
$Z_w(N)\approx A_1~\mu^N ~N^{\gamma_1-1}$ in which the parameter $\mu$
is not affected by the introduction of the membrane, $\gamma_1$ is a
different universal exponent --- $\gamma_1=61/64$ and $\gamma_1\approx
0.68$ in two and three dimensions, respectively --- while $A_1$ is
again not universal. Consider the translocating polymer, for which
there are $n$ monomers on one side and $(N-n)$ monomers on the
other. Since this situation can be seen as two strands of polymers
with one end (of each strand) tethered on the membrane, the number of
states for this polymer is given by $Z_w(n)Z_w(N-n)$, which attains a
minimum when $n=N/2$. The effective entropic barrier faced by a
translocating polymer is thus
\begin{equation}
\Delta S = \log \left( \frac{Z_b(N)}{Z_w(N/2)^2} \right) =c\log(N)+k,
\end{equation}
with $c=\gamma-2\gamma_1+1$ and
$k=\log(A)-2\log(A_1)+2(\gamma_1-1)\log(2)$.

In 2001, Chuang {\it et al.} \cite{kantor} established that the then
existing mean-field type descriptions based on the Fokker-Planck
equation for first-passage over this entropic barrier
\cite{sungpark1,sungpark2,lub} are unsuitable for describing unbiased
translocation. In the absence of explicit hydrodynamics, i.e., for a
Rouse polymer, Chuang {\it et al.\/} \cite{kantor} argued that the
pore-blockade time (or the dwell time) $\tau_d$ cannot be less than
the the Rouse time, which scales as $\tau_R \sim N^{1+2\nu}$. Based on
simulations in 2D using the BFM, with $N=128$ or less, they further
concluded that the ratio of the dwell and Rouse times is approximately
constant, suggesting that $\beta=1+2\nu$ as well. As it later turned
out, the paper by Chuang {\it et al.}  initiated a flurry of
theoretical and simulation works on unbiased translocation.
\begin{table*}
\begin{center}
\begin{tabular}{c|c|c|c|c}
Authors & $\beta$ (2D, Rouse)&$\beta$ (2D, Zimm) & $\beta$ (3D,
Rouse)& $\beta$ (3D, Zimm) \tabularnewline \hline \hline  Chuang {\it
et al.} \cite{kantor}&$1+2\nu=2.5$ (BFM)&---&---&---\tabularnewline
\hline Luo {\it et al.} \cite{luo1} &$2.50\pm0.01$
(BFM)&---&---&---\tabularnewline \hline Huopaniemi {\it et al.}
\cite{luo10} &$2.48\pm0.07$ &---&---&---\tabularnewline  &(FENE
MD)&&&\tabularnewline \hline Wei {\it et al.} \cite{wei}
&$2.51\pm0.03$ &---&$2.2$&---\tabularnewline &(bead-spring
MD)&&(bead-spring MD)&\tabularnewline \hline  Chatelain {\it et al.}
\cite{kantor1}&2.5 (BFM)&---&---&---\tabularnewline \hline Luo {\it et
al.} \cite{luo2} &$2.44\pm0.03$ &---&$2.22\pm0.06$ &---\tabularnewline
& (GROMACS) &---& (GROMACS) &---\tabularnewline \hline Panja {\it et
al.}
\cite{panja1,panja10}&---&---&$2+\nu\approx2.588$&$1+2\nu\approx2.18$\tabularnewline
\hline Panja {\it et al.}
\cite{panja2}&$2+\nu=2.75$&$1+2\nu=2.5$&---&---\tabularnewline \hline
Dubbeldam {\it et al.}
\cite{dub}&---&---&$2.52\pm0.04$&---\tabularnewline  &&&(FENE)
&\tabularnewline \hline Gauthier {\it et al.}
\cite{gary1}&---&---&$2+\nu$ &$1+2\nu$\tabularnewline \hline
Guillouzic {\it et al.}  \cite{gary20}&---&---&---&2.27
(MD)\tabularnewline \hline Gauthier {\it et al.}
\cite{gary2}&---&---&---&11/5=2.2 (MD)\tabularnewline \hline
\end{tabular}
\caption{Summary of all the results on the exponent for the
pore-blockade time for unbiased translocation known to us at the time
of writing this paper. Abbreviations used: MD (molecular dynamics),
FENE (Finite Extension Nonlinear Elastic). \label{table1}}
\end{center}
\end{table*}

For a number of years following the work by Chuang {\it et al.}
\cite{kantor}, several simulation studies reported the exponent for
the pore-blockade time for unbiased translocation both in 2D and 3D to
be consistent with $1+2\nu$ (which in 2D equals 2.5, and $\approx2.18$
in 3D) for a Rouse polymer \cite{luo1,luo10,wei}. Some of these
studies characterized the anomalous dynamics of unbiased translocation
as well: having denoted the monomer number at the pore by $s(t)$ at
time $t$, the mean-square displacement of the monomers $\langle\Delta
s^2(t)\rangle$ through the pore in time $t$ was found to scale $\sim
t^\alpha$ with $\alpha=2/(1+2\nu)$, satisfying the obvious requirement
$\langle\Delta s^2(\tau_d)\rangle=N^2$. Over the last couple of years
however, several other studies on unbiased translocation have been
performed for a Rouse polymer, whose scaling results for $\beta$ and
$\alpha$ differ from $1+2\nu$ and $2/(1+2\nu)$ respectively. In 2007
and 2008, using a theoretical approach based on polymer's memory
effects, aided by simulations with a highly efficient lattice polymer
model (developed by ourselves) we showed that for a Rouse polymer
$\langle\Delta s^2(t)\rangle\sim t^{(1+\nu)/(1+2\nu)}$ [i.e.,
$\alpha=(1+\nu)/(1+2\nu)$] up to the Rouse time $\tau_R$, and
thereafter $\langle\Delta s^2(t)\rangle\sim t$ [i.e., $\alpha=1$] as
no memory in the polymer survive beyond the Rouse time; consequently,
the exponent for the pore-blockade time is given by $\beta=2+\nu$,
i.e., $\approx2.588$ in 3D \cite{panja1,panja10} and $2.75$ in 2D
\cite{panja2}. In the presence of hydrodynamics, i.e., for a Zimm
polymer, $\langle\Delta s^2(t)\rangle$ was predicted to behave $\sim
t^{(1+\nu)/(3\nu)}$ up to the Zimm time $\tau_Z\sim N^{3\nu}$, and
thereafter $\langle\Delta s^2(t)\rangle\sim t$; leading to the
expectation that $\tau_d$ should scale as $N^{1+2\nu}$
\cite{panja1,panja10}. (We showed that the fact that $\beta=1+2\nu$
for a Zimm polymer has nothing to do with Rouse dynamics. It is in
fact a pure coincidence that this exponent is the same as the Rouse
exponent, as explained in Ref. \cite{panja1,panja10}). We showed that
these memory effects stem from the polymer's local strain relaxation
in the neighborhood of the pore \cite{panja1,panja10,panja2,panja3}.

Recent numerical results, using completely different polymer models
from ours, obtained by Dubbeldam {\it et al.} \cite{dub}, and by
Gauthier and Slater \cite{gary1} agree very well with $\beta=2+\nu$
for a Rouse polymer. For a Zimm polymer, Gauthier and Slater
\cite{gary1}, and further works by Guillouzic and Slater
\cite{gary20}, and by Gauthier and Slater \cite{gary2} reported
$\beta=1+2\nu$; these are consistent with our scaling prediction
\cite{panja1,panja10}, but we note that Ref. \cite{gary1} reports this
result using an approach which differs from ours. Thus, while the two
contenders for $\beta$ have emerged to be (a) $1+2\nu$ for a Rouse
polymer, originally proposed by Chuang {\it et al.}  \cite{kantor},
and (b) $2+\nu$ for a Rouse, and $1+2\nu$ for a Zimm polymer,
originally predicted by us \cite{panja1,panja10,panja2}, the
publication of two recent papers \cite{kantor1,luo1} that reassert
their authors' earlier result $\beta=1+2\nu$ for unbiased
translocation for a Rouse polymer indicates that the debate is not yet
settled.

Before proceeding further, for the benefit of the readers, in Table
\ref{table1} we summarize all the results on the exponent for the
pore-blockade time for unbiased translocation known to us to date.

Given that our theoretical arguments for $\beta=2+\nu$ for unbiased
translocation of Rouse polymers are seemingly at odds with a number of
simulations, reporting exponents much closer to $\beta=1+2\nu$, we
decided to redo the latter simulations. It is impossible for us to
analyze in detail each and every model that has been used to produce
$\beta=1+2\nu$ for a Rouse polymer; nevertheless, having seen that the
BFM has been frequently used to obtain this result, we prompt
ourselves to revisit unbiased translocation in 2D for a Rouse polymer,
using {\it exactly the same details\/} of the BFM used by Chuang {\it
et al.}  \cite{kantor}.

First, we extend the range of polymer lengths studied, from $N\le256$
\cite{kantor,kantor1,luo1} up to $N=1000$.  While the reported
behavior for $N\le256$ \cite{kantor,kantor1,luo1} is that the function
$f(N) \equiv \tau_d/N^{1+2\nu}$ is constant within numerical accuracy,
our simulations with longer polymers reveal $f(N)$ to be a
monotonically decreasing quantity with increasing $N$, with a rate of
decrease for $f(N)$ which increases with increasing $N$. The
conclusion is that the reported constant behavior of $f(N)$
corresponds to an {\it effective} exponent of $\beta \approx 1+2\nu$,
which does not hold for long polymers.

Secondly, for the BFM, having established the above for $\tau_d$, we
set out to quantify the finite-size effects in various basic
equilibrium quantities that play a role in the dynamics of
translocation. These are (i) the equilibrium end-to-end distance, (ii)
the (equilibrium) entropic spring constant, and (iii) the longest
correlation time for a tethered polymer at equilibrium as a function
of their length. We find, for the BFM, that the finite-size effects
for (i) are negligible (data not shown in this paper), but the
finite-size effects for (ii) and (iii) are severe. Once these
finite-size effects are taken into account, the polymer's memory
effects for the BFM do not rule out those we found in
Refs. \cite{panja1,panja10}, which originally reported $\beta=2+\nu$.

Thirdly, and quite remarkably, these finite-size effects still do not
explain the peculiar behaviour of $f(N)$ for the BFM.  We therefore
also study a dynamic quantity, namely the mobility of individual
monomers in the BFM, as a function of monomer number and direction.
Especially for the monomers in the vicinity of the pore, this dynamic
quantity shows highly anomalous behavior.  We find that when this
behavior is corrected towards how one expects the monomers to behave
in the neighborhood of the pore, the exponent for the pore-blockade
time increases towards $2+\nu$. We therefore conclude that the BFM is
a fine model for polymer dynamics in general, but does not handle
situations very well where the polymer is constrained to pass through
a narrow pore. Our analysis also implies that for those polymer models
that assert $\beta=1+2\nu$, one needs to thoroughly investigate their
finite-size effects and dynamical peculiarities, if such an assertion
is to prove meaningful.

The organization of this paper as follows. In Sec. \ref{sec2} we
briefly discuss the BFM, and the details of the used BFM variant in
this paper. In Sec. \ref{sec3} we study finite-size effects in the BFM
for several equilibrium properties.  In Sec. \ref{sec4} we study with
the BFM the quantity $f(N)=\tau_d/N^{1+2\nu}$, and find that it is a
monotonically decreasing quantity with increasing $N$, such that the
rate of decrease for $f(N)$ increases with increasing $N$; this result
implies that the reported exponent $\beta=1+2\nu$ for Rouse polymers
is an {\it effective} exponent which does not hold in the limit
$N\rightarrow\infty$, contrary to the conclusions of
Refs. \cite{kantor,kantor1,luo1,luo2}. In Sec. \ref{sec4} we also show
that we do not find $\alpha=2/(1+2\nu)$ [$=0.8$ in 2D], while we do
find a diffusive regime for the anomalous dynamics of translocation,
contradicting the claims of Ref. \cite{luo1}. In Sec. \ref{sec4a} we
demonstrate that once the finite-size effects in various basic
quantities for the BFM are taken into account, the polymer's memory
effects for the BFM are entirely consistent with those we found in
Refs. \cite{panja1,panja10}, which originally reported
$\beta=2+\nu$. In Sec. \ref{sec5} we trace the peculiarities of the
BFM to the anomalous behavior of the mobility of individual monomers
in the neighborhood of the pore. We conclude the paper in
Sec. \ref{sec6}.

All throughout this paper, following the convention of the existing
literature, we denote the monomer number located in the pore at time
$t$ by $s(t)$.

\section{The bond fluctuation model (BFM)\label{sec2}}

\begin{figure}[!h] 
\includegraphics[width=0.8\columnwidth]{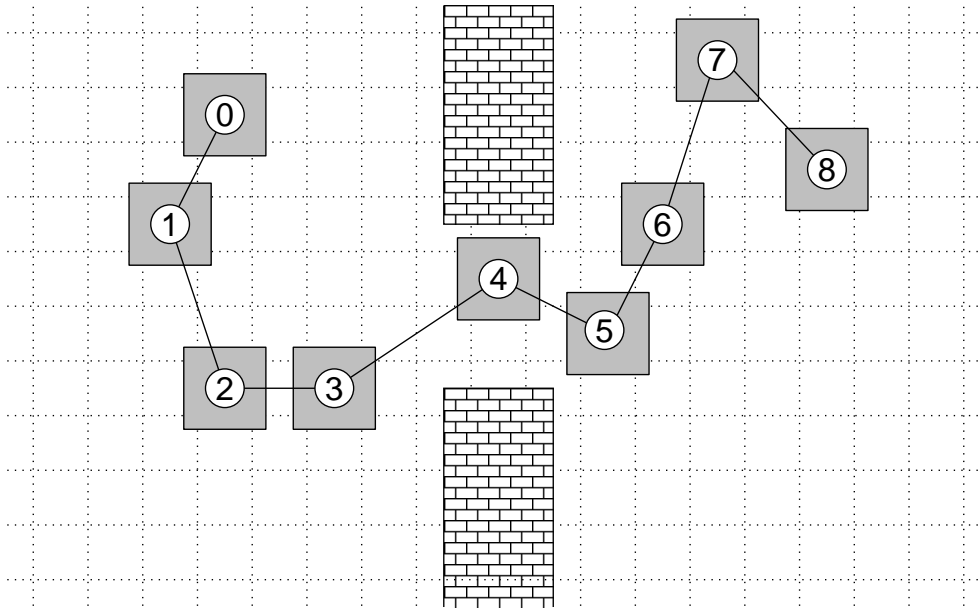}
\caption{The original bond fluctuation model in 2D, due to Carmesin
and Kremer \cite{carm}, illustrated for a polymer translocating
through a membrane thickness of two lattice sites thickness, with
monomer 4 residing within the pore that is three lattice sites
wide. Monomers reside on a square lattice. Adjacent monomers are
connected by a bond, which can only take lengths of $2$, $\sqrt{5}$,
$2\sqrt{2}$, $3$, $\sqrt{10}$, or $\sqrt{13}$. The excluded area
around each monomer consists of $2\times 2$ lattice sites. The
particular choice for the values allowed to the bond length, in
combination with the size of the excluded area, suffices to avoid the
crossing of bonds. In simulations of translocation, the membrane is
represented as a line with a thickness of 2 sites, with a pore of 3
lattice sites wide.  The dynamics of the polymer consists of
single-monomer hops to nearest-neighbor lattice sites, restricted by
the constraints on bond length and excluded volume.\label{figbond2d}}
\end{figure}

The two-dimensional bond fluctuation model, introduced by Carmesin and
Kremer in 1988 \cite{carm}, is a very frequently used model for the
simulation of polymer dynamics. In the original form of the
model, each monomer occupies four ($2 \times 2$) lattice sites of a
square lattice: thus two monomers are always separated by at least a
distance of two lattice spacings. Monomers adjacent in the polymer are
connected by bonds with lengths between 2 and $\sqrt{13}$. The
original model is illustrated for a polymer through a membrane of
thickness two lattice sites in Fig. \ref{figbond2d}, with monomer 4
residing within the pore that is three lattice sites wide.

As mentioned already, in this paper we use exactly the same model as
that of Ref. \cite{kantor}, which is a variation on the original BFM
due to Carmesin and Kremer. Its details are as follows. Hydrodynamic
interactions are not considered in this model. The $N$ monomers of the
polymer reside on a square lattice. Excluded volume interactions are
implemented by forbidding two monomers to be closer than $2$ lattice
units, while the sequential connectivity of the monomers are
maintained by requiring the separation between the adjacent monomers
of the polymer to be less than or equal to $\sqrt{10}$ lattice
units. This choice of the minimal and the maximal distances ensures
that the polymer never intersects itself. The dynamics of the model is
implemented by Monte Carlo moves: An elementary move consists of an
attempt to move a randomly selected monomer by one lattice spacing in
an arbitrarily chosen direction. If the new configuration is
permitted, the move is accepted; otherwise, the move is rejected. The
unit of time in this model is defined by $N$ attempted MC moves for
the entire polymer. We choose a box size of $10N\times10N$; and the
membrane, with a thickness of two lattice units, divides the box into
two equal chambers of size $10N\times5N$, with the pore of width three
lattice units exactly at the center of the box. The tight size of the
pore ensures that the monomer $s$ residing within the pore is uniquely
defined at any time. For each realization, we first tether the polymer
halfway at the pore, with $N/2$ monomers on each side of the membrane,
and equilibrate it for times typically $>100N^{1+2\nu}$. We then
remove the tether at time $t=0$ and wait till the polymer disengages
from the pore. The averages are obtained over an ensemble of such
realizations: we use $16,384$ realizations for $N<1000$ and $2,048$
realizations for $N=1000$. We define the mean time $\tau_u$ that the
polymer takes to disengage from the pore to either side of the
membrane as the characteristic unthreading time for the polymer. For
unbiased translocation, the scaling of the unthreading time with
polymer length is the same as that of $\tau_d$ \cite{panja10}.
\begin{table*}
\begin{center}
\begin{tabular}{p{2cm}|p{2cm}|p{1.5cm}|p{1.5cm}|p{3.5cm}}
\hspace{8mm}$N$ & \hspace{5mm}$\langle R_e\rangle$&
\hspace{5mm}$\langle R^2_e\rangle$& \hspace{6mm}${\cal R}^2$&
\hspace{5mm}${\cal R}^2/(N/2)^{2\nu}$ \tabularnewline \hline \hline
\hspace{7mm}20 & \hspace{4mm}13.60&
\hspace{4mm}197.8& \hspace{4mm}12.90& \hspace{9mm}0.408
\tabularnewline \hline \hspace{7mm}30 & \hspace{4mm}18.39&
\hspace{4mm}363.7& \hspace{4mm}25.40& \hspace{9mm}0.437
\tabularnewline \hline \hspace{7mm}40 & \hspace{4mm}22.81&
\hspace{4mm}560.9& \hspace{4mm}40.74& \hspace{9mm}0.455
\tabularnewline \hline \hspace{7mm}50 & \hspace{4mm}26.92&
\hspace{4mm}783.2& \hspace{4mm}58.66& \hspace{9mm}0.469
\tabularnewline \hline \hspace{7mm}60 & \hspace{4mm}30.83&
\hspace{4mm}1029.5& \hspace{4mm}78.83& \hspace{9mm}0.480
\tabularnewline \hline \hspace{7mm}80 & \hspace{4mm}38.20&
\hspace{4mm}1583.7& \hspace{4mm}124.5& \hspace{9mm}0.492
\tabularnewline \hline \hspace{7mm}100 & \hspace{4mm}45.13&
\hspace{4mm}2213.5& \hspace{4mm}176.9& \hspace{9mm}0.500
\tabularnewline \hline \hspace{7mm}150 & \hspace{4mm}61.01&
\hspace{4mm}4056.3& \hspace{4mm}334.5& \hspace{9mm}0.515
\tabularnewline \hline \hspace{7mm}200 & \hspace{4mm}75.67&
\hspace{4mm}6247.0& \hspace{4mm}520.3& \hspace{9mm}0.520
\tabularnewline \hline \hspace{7mm}300 & \hspace{4mm}102.4&
\hspace{4mm}11455& \hspace{4mm}969.4& \hspace{9mm}0.528
\tabularnewline \hline \hspace{7mm}500 & \hspace{4mm}149.9&
\hspace{4mm}24584& \hspace{4mm}2108&
\hspace{9mm}0.533 \tabularnewline \hline \hspace{7mm}700 &
\hspace{4mm}193.0& \hspace{4mm}40751&
\hspace{4mm}3513& \hspace{9mm}0.537 \tabularnewline \hline
\hspace{7mm}1000 & \hspace{4mm}252.2& \hspace{4mm}69616&
\hspace{4mm}6018& \hspace{9mm}0.538 \tabularnewline \hline
\end{tabular}
\caption{Scaling of the inverse entropic spring constant ${\cal R}^2$ of
the polymer, data averaged over $10^7$ realizations, which are separated
by one pivot move and N BFM moves; this corresponds roughly to a million
statistically independent measurements.  The systematic trend as shown
in 5th column demonstrates the strong finite-size effects in the BFM.
\label{table2}}
\end{center}
\end{table*}

\section{Equilibrium properties for the BFM\label{sec3}}

Two main ingredients for the derivation of \hbox{$\beta=2+\nu$} for a
Rouse polymer, as predicted by us \cite{panja1,panja10,panja2}, are
the following well-known properties of Rouse polymers. For a polymer
of length $N$ with one end tethered on a membrane, (i) the inverse
entropic spring constant should scale as $N^{2\nu}$ \cite{degennes},
and (ii) the equilibrium correlation function for the tether-to-end
vector must scale as $N^{1+2\nu}$. We now check for both
properties.  
\begin{figure*}
\begin{center}
\begin{minipage}{0.45\linewidth} 
\includegraphics[width=0.8\linewidth,angle=270]{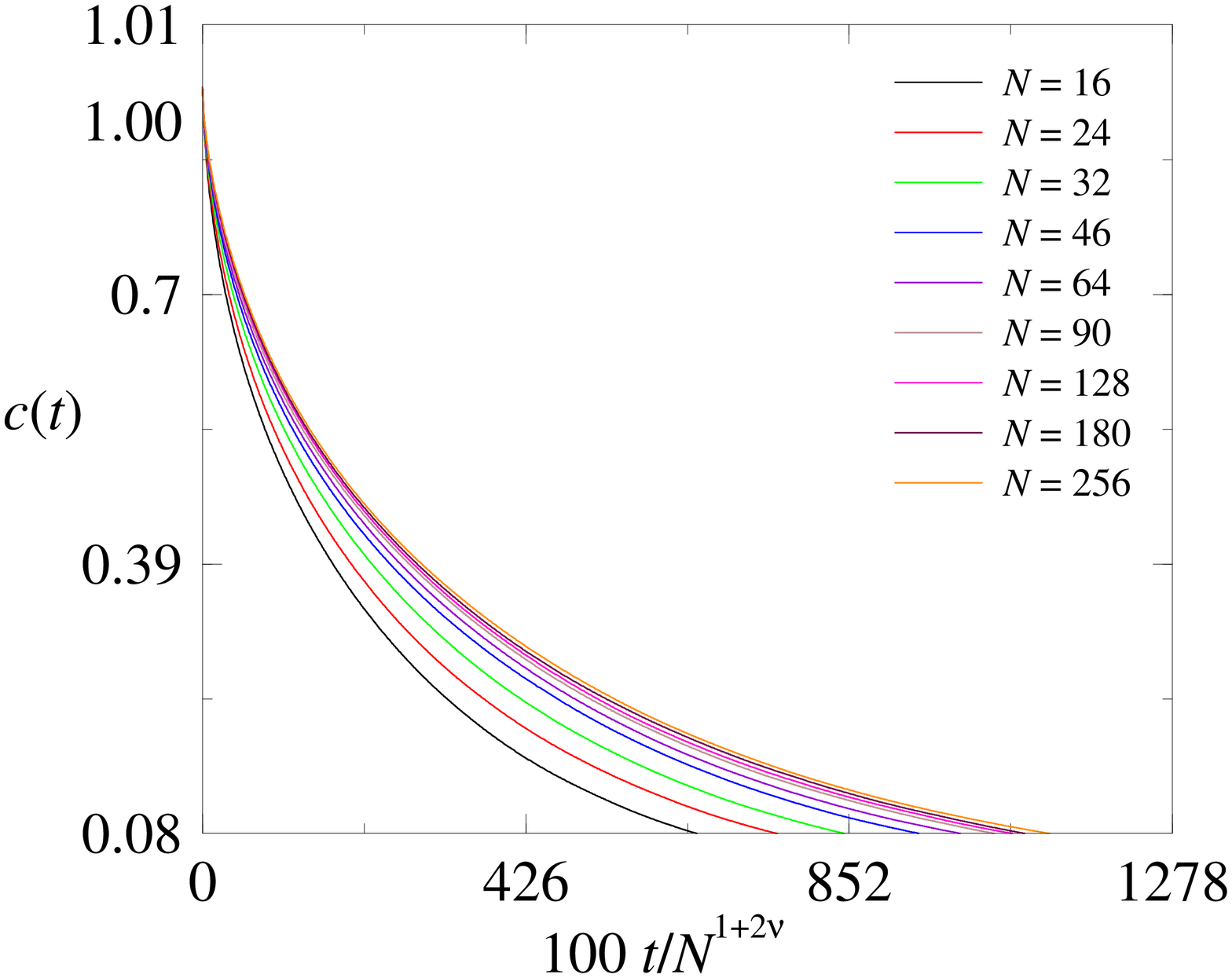}
\end{minipage} \hspace{10mm}
\begin{minipage}{0.45\linewidth}
\includegraphics[width=0.8\linewidth,angle=270]{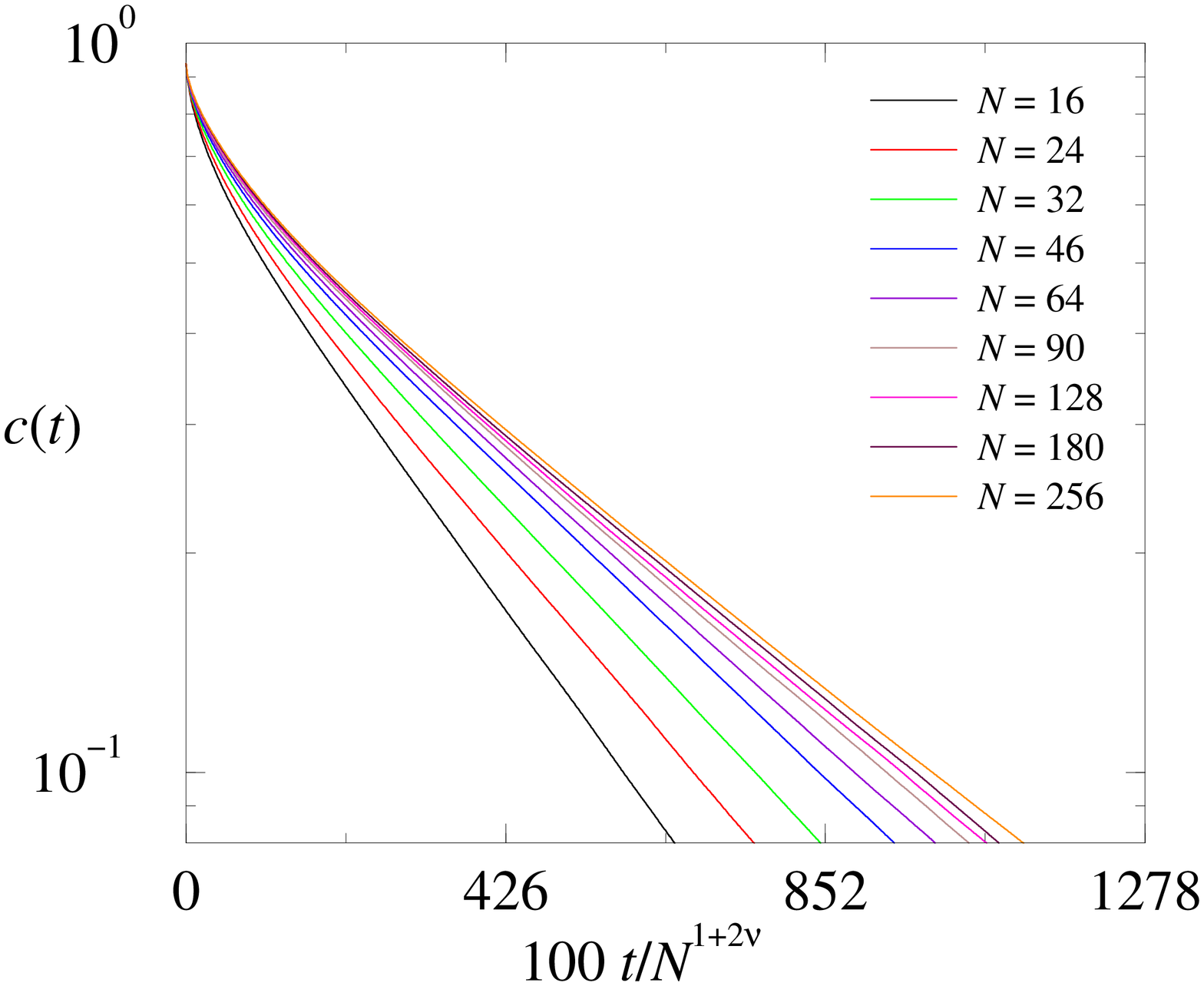}
\end{minipage} 
\end{center}
\caption{Plot of the correlation function $c(t)$ for
different values of $N$: linear scale (left panel), semi-log scale
(right panel), with $\nu=0.75$ in 2D. The lack of collapse
demonstrates the presence of strong finite-size effects in the
dynamical properties of polymers in BFM. The $N=256$ data corresponds
well to $c_{(N=256)}(t)\sim\exp[-t/(5.8\times10^6)]$. Data averaged
over 16,384 independent realizations.
\label{fig1}}
\end{figure*}

\subsection{Entropic spring constant of the polymer\label{sec3a}}

With the polymer threaded halfway at the pore, i.e., effectively for a
polymer of length $N/2$ with one end tethered on the membrane, we
denote the distance of the free ends of the polymer from the pore by
$R_e$, and then the inverse of the spring constant of the polymer is
$\propto{\cal R}^2=\langle R^2_e\rangle-\langle R_e\rangle^2$
\cite{degennes}, where the angular brackets denote the average over
the equilibrium ensemble --- equilibrium is achieved by applying a
million pivot moves to each realization upon tethering one end of the
polymer on the membrane. For long polymers the inverse entropic
spring constant ${\cal R}^2$ should scale as $N^{2\nu}$. Since this
quantity is an equilibrium property, we combined the usual
single-monomer moves of the BFM with pivot moves, in which rotations
of the polymer tails by $\pm$90 degrees around a randomly selected
monomer are attempted, and accepted if the resulting configuration is
valid. With pivot moves, care has to be taken that not only
overlapping monomers cause rejection: also attempted moves to other
configurations which are not accessible via a sequence of the usual
single-monomer moves should be rejected. Because of the fast
decorrelation of the combined algorithm, accurate measurements could
be obtained. The finite-size effects in the scaling of ${\cal R}^2$
are shown in Table \ref{table2}.

\subsection{Equilibrium correlation function for the tether-to-end
  vector for the BFM\label{sec3b}}

Similarly, for a polymer of length $N/2$ with one end tethered on the
membrane, we denote the vector distance of the free end of the polymer
w.r.t. the tethered end at time $t$ by $\mathbf{e}(t)$, and define the
correlation function for the tether-to-end vector as
\begin{eqnarray} c(t) =
\frac{\langle\mathbf{e}(t)\cdot\mathbf{e}(0)\rangle-\langle\mathbf{e}(t)\rangle\cdot\langle\mathbf{e}(0)\rangle}{\sqrt{\langle\mathbf{e}^2(t)-\langle\mathbf{e}(t)\rangle^2\rangle\langle\mathbf{e}^2(0)-\langle\mathbf{e}(0)\rangle^2\rangle}}\,.
\label{e1}
\end{eqnarray} 
The angular brackets in Eq. (\ref{e1}) denote averaging in
equilibrium. Here equilibrium means that with one end tethered on the
membrane the polymer is thermalized for times $>100N^{1+2\nu}$. The
quantity $c(t)$ is plotted in Fig. \ref{fig1} for several values of
$N\le256$, in linear as well as in semi-log plots. The lack of
collapse in Fig. \ref{fig1}, together with the data of Table
\ref{table1}, demonstrates that the finite-size effects in this model
are severe.

\section{Asymptotic scaling of $\tau_d$, and anomalous dynamics of
unbiased translocation in the BFM\label{sec4}}

\subsection{In the BFM the scaling $\tau_d\sim N^{1+2\nu}$ does not
  hold asymptotically\label{sec4a}}

Having shown that the finite-size effects of the BFM are significant
at least up to lengths of a few hundred monomers, below in Table
\ref{table3} we present the results of the mean unthreading time
$\tau_u$ over a wide range of values of $N$. From Table \ref{table3}
we find that for the BFM the quantity $f(N)=\tau_d/N^{1+2\nu}$ is a
monotonically decreasing quantity with increasing $N$, such that the
rate of decrease for $f(N)$ increases with increasing $N$. Having
noted that for the BFM, the conclusion that $\beta=1+2\nu$ has been
based on simulation data for $N\le256$
\cite{kantor,kantor1,luo1,luo2}, the finite-size effects in this model
as demonstrated in Sec. \ref{sec3} and the data of Table \ref{table3}
imply that there is no convincing numerical evidence that the exponent
$\beta$ approaches the value $1+2\nu$ in the thermodynamic limit
($N\rightarrow\infty$), contrary to the conclusions of
Refs. \cite{kantor,kantor1,luo1,luo2}. Rather, it is an {\it
effective} exponent, approximately valid over a finite range of
polymer lengths.  It is worthwhile to mention here that in contrast to
the behavior of $\beta$ obtained from the BFM, our result
$\beta=2+\nu$ for a Rouse polymer has been checked for $N$ up to $500$
\cite{panja1,panja10}, and up to $1000$ \cite{henkthesis} in our
lattice polymer model, and we have shown that in our model the
finite-size effects become undetectable beyond $N=150$ \cite{panja10}.
\begin{figure*}
\begin{center}
\begin{minipage}{0.49\linewidth} 
\includegraphics[angle=270,width=\linewidth]{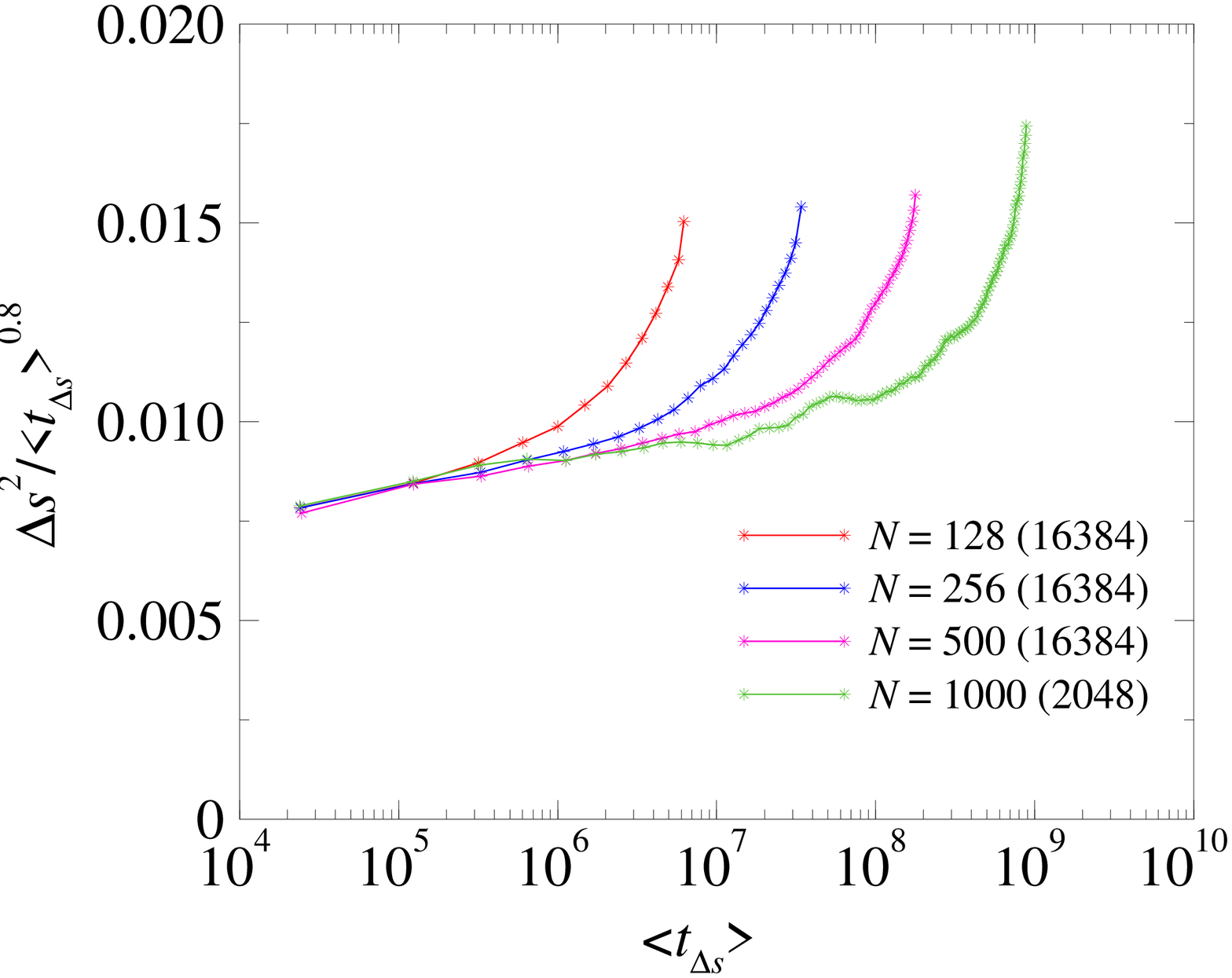}
\end{minipage} \hspace{-2mm}
\begin{minipage}{0.49\linewidth}
\includegraphics[angle=270,width=\linewidth]{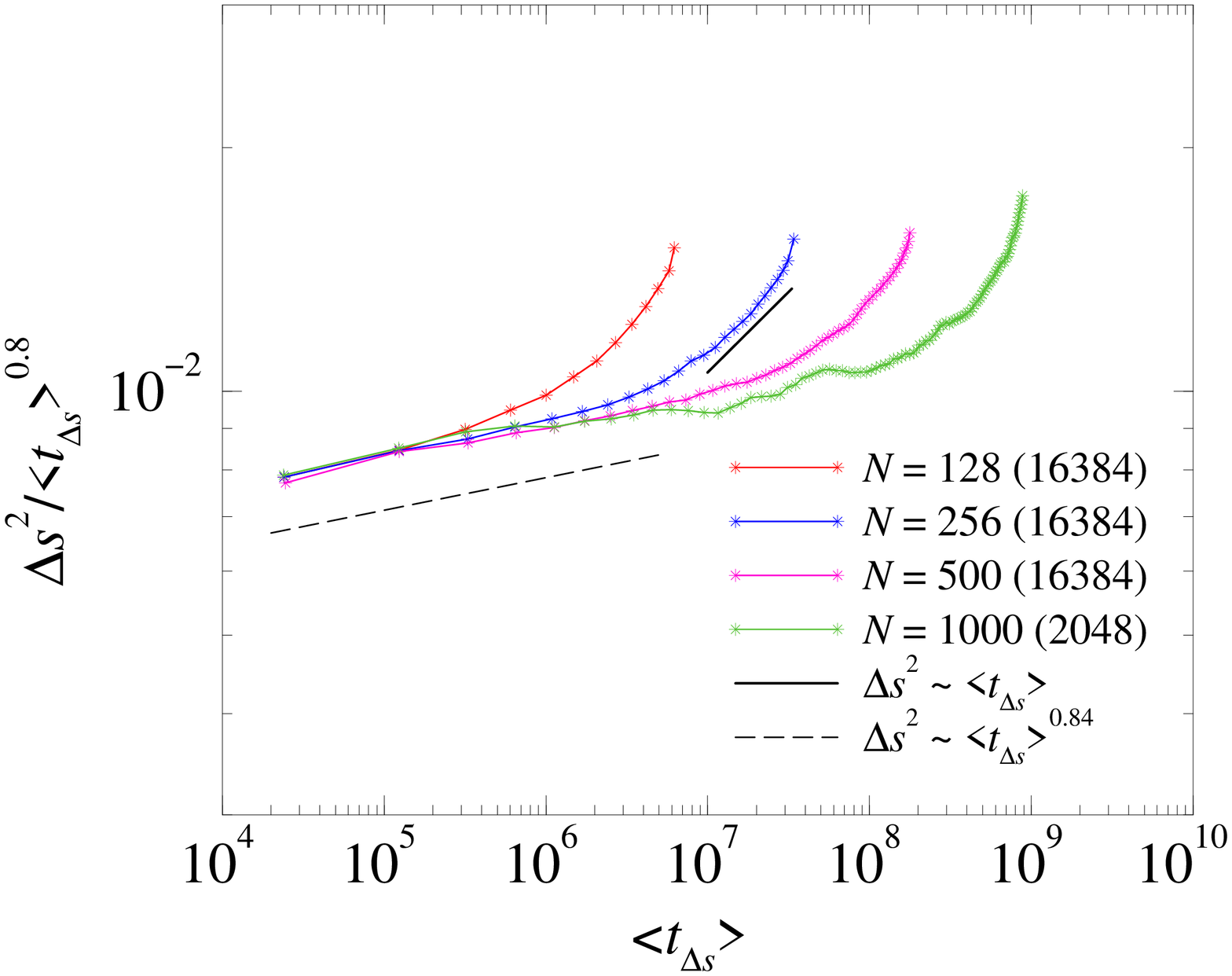}
\end{minipage} 
\end{center}
\caption{Plots of $\Delta s^2/\langle t_{\Delta
s}\rangle^{0.8}$ as a function of $\langle t_{\Delta s}\rangle$ with
$|\Delta s|=5,10,15,\ldots,N/2$: semi-log (left panel), log-log (right
panel). The numbers in the parentheses in the legends denote the
number of independent realizations over which the averages have been
calculated. If indeed $\langle\Delta s^2(t)\rangle$ were to scale as
$t^{2/(1+2\nu)}=t^{0.8}$, both panels would have produced horizontal
lines. The lowest exponent $\alpha$ for $\langle\Delta
s^2(t)\rangle\sim t^{\alpha}$ obtained for each curve is $0.84$, which
gradually crosses over to diffusive regime (for $|\Delta s|$  close to
$N/2$, because of entropic reasons, $\alpha>1$); at a contrast to
this, $\alpha=0.86$ has been reported in Ref. \cite{kantor1} all
throughout; implying that the value of $\alpha$ reported in
Ref. \cite{kantor1} is an effective value. Additionally there is a
diffusive regime, contrary to the claims of Ref. \cite{luo2}, as shown
by the solid black line in the right panel. As explained in the text,
this diffusive regime is to be expected since the mean unthreading time is
much longer than the longest relaxation times of the polymers. The
plots are clearly consistent with the data in Table \ref{table2},
namely that the coefficient $\tau_u/N^{1+2\nu}$ is a decreasing
function of $N$. \label{fig3}}
\end{figure*}

\subsection{Anomalous dynamics of unbiased translocation in the
  BFM\label{sec4b}}

Next we focus on the anomalous dynamics of unbiased translocation in
the BFM. The standard way to study the anomalous dynamics is to plot
$\langle\Delta s^2(t)\rangle$ as a function of $t$, as discussed in
the introduction; however, it comes with the following
disadvantage. With the initial condition $s(t=0)=N/2$, $\Delta s$ is
undefined once the polymer disengages from the pore, and as a result,
$\langle\Delta s^2(t)\rangle$ saturates in time fairly quickly
\cite{kantor1}. To avoid such saturation effects, we study the
anomalous dynamics of translocation in a somewhat non-standard
form. We calculate the mean first passage time $\langle t_{\Delta
s}\rangle$ of arrival at a monomeric distance $\Delta s \equiv \left|
s(t)-N/2 \right|$.  Thus, in each simulation, starting with $s=N/2$ at
$t=0$, we keep track of the first arrival times $t_{\Delta s}$ for all
values of $\Delta s = 1\dots N/2$.  In this manner, at no point in
time the data suffer from saturation problems.  The idea behind
obtaining $t_{\Delta s}$ as a function of $\Delta s$ is actually
motivated by the observation that $\tau_u$ is the mean first passage
time for $\Delta s=N/2$.

To verify whether the purported scaling $\langle \Delta
s^2(t)\rangle\sim t^\alpha$ with $\alpha=2/(1+2\nu)$
\cite{kantor,luo1,kantor1} holds, we plot $\Delta s^2/\langle
t_{\Delta s}\rangle^{0.8}$ in Fig. \ref{fig3}  as a function of
$\langle t_{\Delta s}\rangle$ (left panel: log-linear, right panel:
log-log). The data should then be constant so long as the polymer
remains threaded within the pore, since $1+2\nu=0.8$ in 2D. However,
as Fig. \ref{fig3} demonstrates, we do not find any evidence for
$\alpha=2/(1+2\nu)=0.8$; the lowest value of $\alpha$ we find is
$0.84$ (dashed line in the right panel of Fig. \ref{fig3}), which
slowly crosses over to diffusive behavior at long times [contrary to
the report of Ref. \cite{luo1}, where, based on $\langle \Delta
s^2(t)\rangle$ data till $t\lesssim10^3$, it has been concluded that
there is no diffusive regime for the dynamics of unbiased
translocation]. In this context, note that close to $|\Delta s|=N/2$,
the effective exponent $\alpha$ exceeds 1, as it should be, since
close to the end, an entropic driving force takes over, giving rise to
a non-negligible velocity of translocation. The diffusive behavior is
illustrated in the right panel of Fig. \ref{fig3} for $N=256$; the
existence of the diffusive regime is expected to show beyond the
autocorrelation time, which for a polymer of length $N=256$ equals
$\tau_c\approx 5.8\times 10^6$ (see Fig. \ref{fig1}), but well below
the unthreading time $\tau_u^{(N=256)}\approx3.8\times10^7$. Also note
here that $\alpha>0.8$ is consistent with the observation that
$f(N)=\tau_u/N^{1+2\nu}$ is a monotonically decreasing function of
$N$, as shown in Table \ref{table3}.
\begin{table}[h]
\begin{center}
\begin{tabular}{p{2cm}|p{4.1cm}|p{3.5cm}} \hspace{8mm}$N$ &
\hspace{20mm}$\tau_u$ &$\,\,\gamma(N)=\tau_u/N^{1+2\nu}$
\tabularnewline \hline \hline \hspace{7mm}16 &
\hspace{6mm}$(3.53\pm0.06)\times 10^4$ &
\hspace{9mm}$34.5\pm0.6$ \tabularnewline \hline \hspace{7mm}24 &
\hspace{5mm}$(1.01\pm0.02)\times 10^5$& \hspace{9mm}$35.6\pm0.6$
\tabularnewline \hline
\hspace{7mm}32 & \hspace{5mm}$(2.04\pm0.04)\times 10^5$ &
\hspace{9mm}$35.1\pm0.6$ \tabularnewline \hline \hspace{7mm}46 &
\hspace{5mm}$(4.96\pm0.09)\times 10^5$ &
\hspace{9mm}$34.6\pm0.6$ \tabularnewline \hline \hspace{7mm}64 &
\hspace{4mm}$(1.10\pm0.02)\times 10^6$&
\hspace{9mm}$33.7\pm0.6$\tabularnewline \hline
\hspace{7mm}90 & \hspace{4mm}$(2.56\pm0.04)\times 10^6$&
\hspace{9mm}$33.3\pm0.6$\tabularnewline \hline \hspace{7mm}128 &
\hspace{4mm}$(6.1\pm0.1)\times 10^6$&
\hspace{9mm}$33.2\pm0.6$\tabularnewline \hline
\hspace{7mm}180 & \hspace{4mm}$(1.44\pm0.02)\times 10^7$&
\hspace{9mm}$33.1\pm0.6$\tabularnewline \hline \hspace{7mm}256 &
\hspace{4mm}$(3.38\pm0.06)\times 10^7$&
\hspace{9mm}$32.2\pm0.6$\tabularnewline \hline
\hspace{7mm}500 & \hspace{4mm}$(1.73\pm0.03)\times 10^8$&
\hspace{9mm}$31.0\pm0.6$\tabularnewline \hline \hspace{7mm}1000 &
\hspace{4mm}$(8.79\pm0.01)\times 10^8$&
\hspace{9mm}$27.9\pm0.4$\tabularnewline \hline
\end{tabular}
\caption{Mean unthreading time over 2,048 runs for each value of
$N$.\label{table3}}
\end{center}
\end{table}

\section{Polymer's memory effects: the theory of translocation and the
  BFM\label{sec4c}}

In Refs. \cite{panja1,panja10} we presented the theory of
translocation, based on the polymer's memory effects. We now
demonstrate that once these finite-size effects are taken into
account, the polymer's memory effects for the BFM are consistent with
that theory.

The theory we presented in Refs. \cite{panja1,panja10} is as
follows. Translocation takes place via the exchange of monomers
through the  pore. This exchange responds to $\phi(t)$, the difference
in chain tension perpendicular to the membrane; simultaneously,
$\phi(t)$ adjusts to $v(t)=\dot{s}(t)$, the transport velocity of
monomers across the pore, as well. In the presence of memory effects,
$\phi(t)$ and $v(t)$ are related to each other by
$\phi(t)=\int_{0}^{t}dt' \mu(t-t')v(t')$ via the memory kernel
$\mu(t)$. This relation can be inverted to obtain
$v(t)=\int_{0}^{t}dt' a(t-t')\phi(t')$. The uniqueness of the relation
between $\phi(t)$ and $v(t)$ implies that in the Laplace transform
language, $\mu(k)=a^{-1}(k)$, where $k$ is the Laplace variable
representing inverse time. Additionally, via the
fluctuation-dissipation theorem, $\mu(t-t')$ and $a(t-t')$ are
expressed as
\begin{eqnarray}
\mu(t-t')\!=\!\langle\phi(t)\phi(t')\rangle_{v=0};\,\,
a(t-t')\!=\!\langle v(t)v(t')\rangle_{\phi=0}.
\label{eqnew}
\end{eqnarray}
\begin{figure*}
\begin{minipage}{0.49\linewidth}
\includegraphics[width=\linewidth]{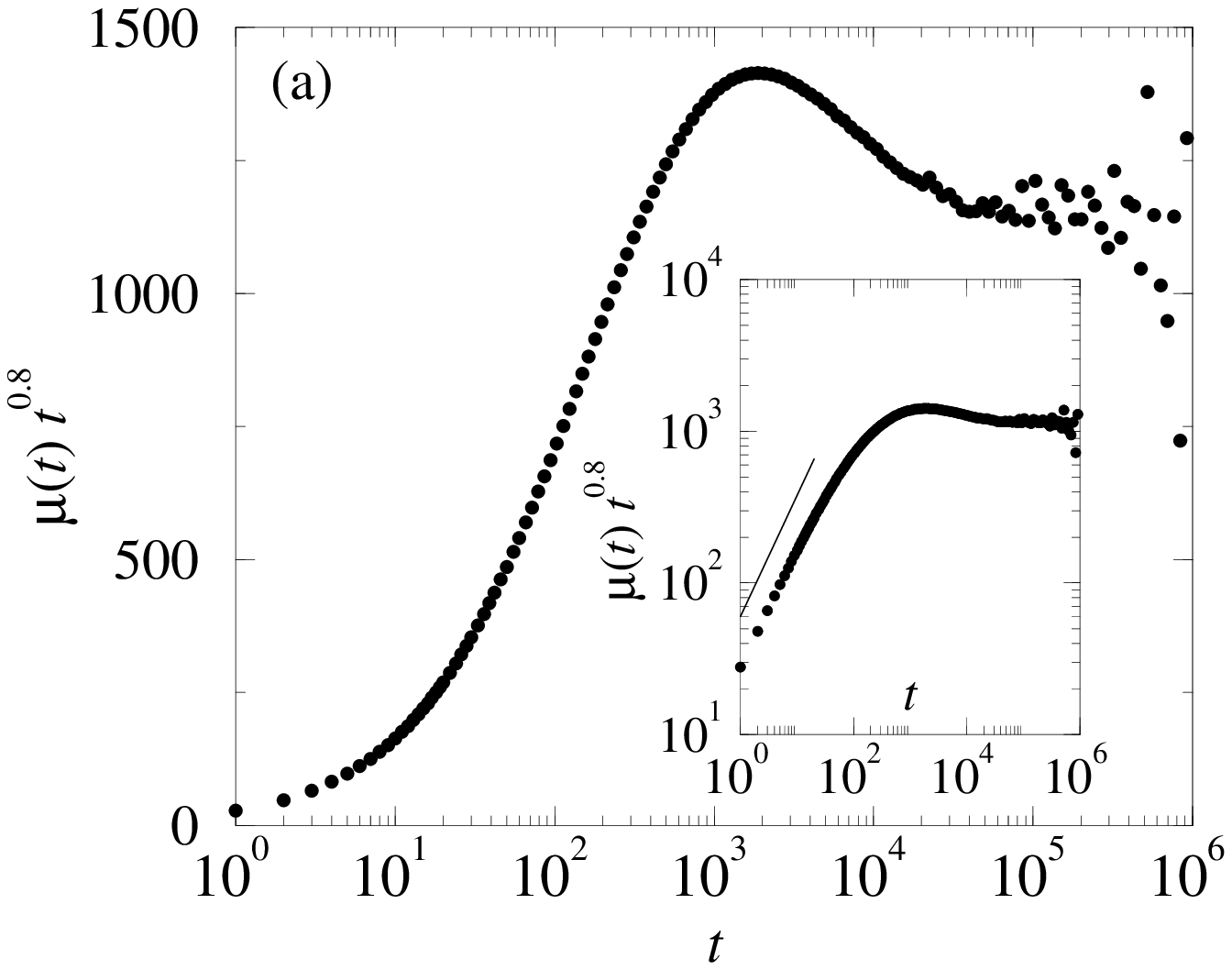}
\end{minipage}
\begin{minipage}{0.49\linewidth}
\includegraphics[width=\linewidth]{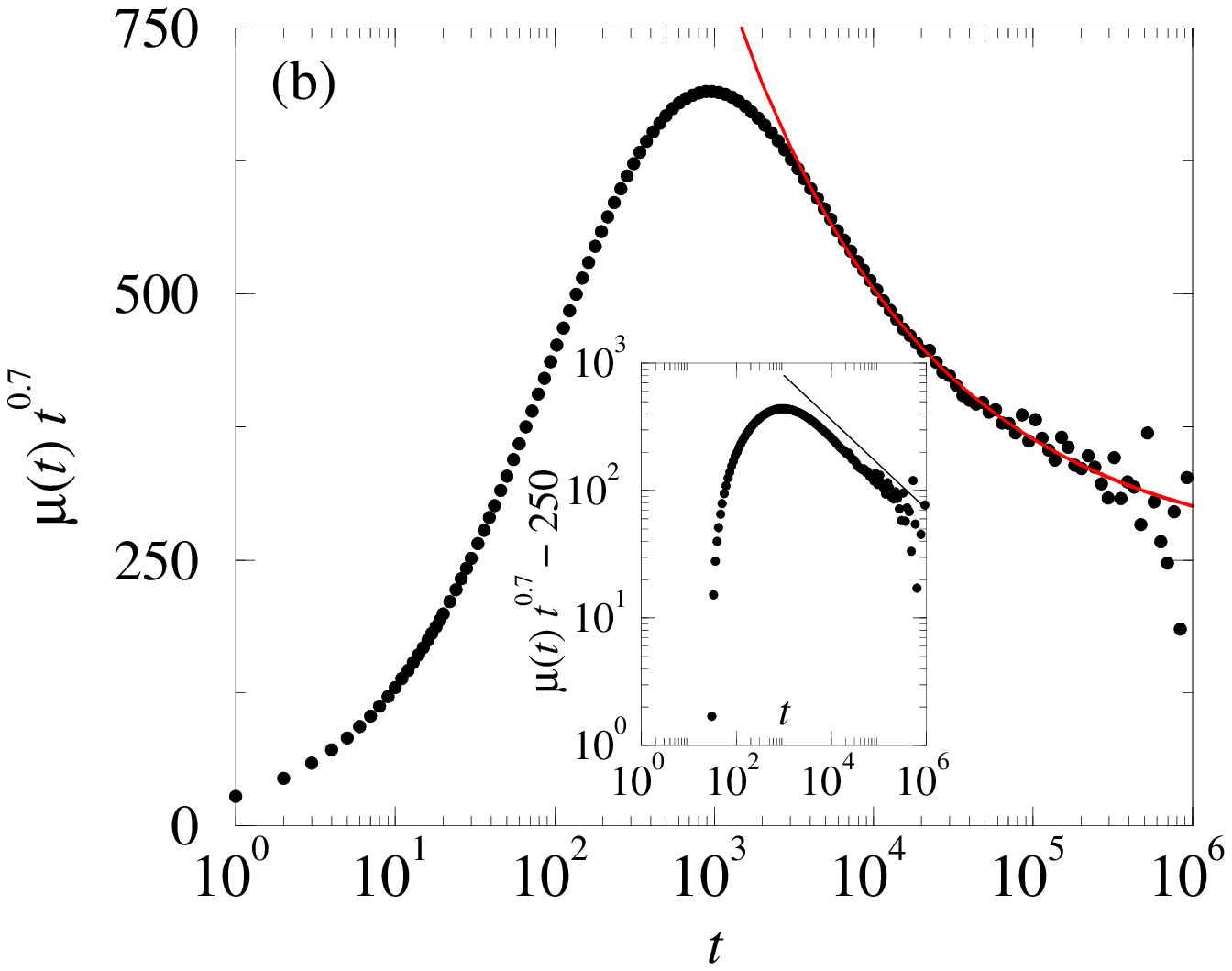}
\end{minipage}
\caption{The memory kernel $\mu(t)$ for a polymer with length $N=1000$
for the BFM. (a) Demonstrates the lack of a single power-law $\sim
t^{-0.8}$, (main graph: semi-log, inset: same data shown in
log-log). The initial increasing part is caused by $\mu(t)$ remaining
practically constant till $t\approx 500$, this is shown by the solid
line in the inset corresponding to an exponent $0.8$. (b) However,
given that we expect a behavior $\mu(t)\sim
t^{-(1+\nu)/(1+2\nu)}=t^{-0.7}$ \cite{panja1,panja10} for the BFM in
2D, we show that the late-time data can be numerically well-fitted by
$6400t^{-1.05}+250t^{-0.7}$, as shown by the red line. The inset
demonstrates the existence of a second power-law behavior of $\mu(t)$
at intermediate times: the solid line corresponds to the scaling
$\sim t^{-0.35}$. \label{fignew1}}
\end{figure*} 

In Refs. \cite{panja1,panja10} we showed that the polymer's memory
kernel is given by $\mu(t)\sim
t^{-\frac{1+\nu}{1+2\nu}}\exp(-t/\tau_R)$, in which $\tau_R$ is the
Rouse time; this result, together with Eq. (\ref{eqnew}) yields
$\tau_d\sim N^{2+\nu}$. The derivation for the exponent
$(1+\nu)/(1+2\nu)$ of the power-law relies on three scaling relations
for an equilibrated polymer of length $n$ with one end tethered to a
membrane: (i) the real-space distance between the free and the
tethered end scales as $n^\nu$, (ii) its entropic spring constant
scales as $n^{-2\nu}$, and (iii) its longest correlation time scales
as $n^{1+2\nu}$. For the BFM, (i) holds for $n$ not so large, but
since the scalings (ii) and (iii) suffer from severe finite-size
effects (as reported in Secs. \ref{sec3a} and \ref{sec3b}), we expect
the $t^{-\frac{1+\nu}{1+2\nu}}$ behavior of the power-law in $\mu(t)$
to only manifest itself at long times. Note that ``long times'' here
refers to times long compared to unity, but short in comparison to the
longest relaxation time of the polymer; this implies that $\mu(t)\sim
t^{-\frac{1+\nu}{1+2\nu}}$ can only be observed when the polymer is
long. Indeed, we demonstrate this below in Fig. \ref{fignew1}, by
measuring $\mu(t-t')=\langle\phi(t)\phi(t')\rangle_{v=0}$ for the BFM
for a polymer with length $N=1000$, where we used the
perpendicular-to-the-membrane distance $Z_4$ of the center-of-mass of
the first four monomers (counting from the pore) as a proxy for the
chain tension \cite{comment}.  More precisely, we tether the middle
monomer of the polymer in the pore (this corresponds to $v=0$), and
obtain good statistics for $\langle\phi(t)\phi(t')\rangle$ over a
total simulation time of $4\cdot 10^{10}$ attempted moves per
monomer. As can be seen in Fig. \ref{fignew1}(a),
$\mu(t)=\langle\phi(t)\phi(0)\rangle_{v=0}$ does not show a memory
exponent $0.8$. Moreover, given that we expect a behavior $\mu(t)\sim
t^{-(1+\nu)/(1+2\nu)}=t^{-0.7}$ \cite{panja1,panja10} for the BFM in
2D, Fig. \ref{fignew1}(b) shows that $\mu(t)$ can be fitted with a
combination of power-laws $\sim t^{-0.7}$ and $\sim t^{-1.05}$ (see
figure caption for details).  The fact that the data can be fitted
with this combination of power-laws does not constitute compelling
evidence for $\mu(t)\sim t^{-0.7}$; other exponents within a range
of $\pm 0.1$ can be fitted as well, with suitably chosen power-law
corrections. The main point is, however, that our theoretically
expected behavior $\mu(t)\sim t^{-0.7}$ cannot be ruled out from these
numerical data.

\section{Anomalous monomeric mobility for the BFM in and near the pore
  \label{sec5}} 

So far we have discussed the finite-size effects in several basic
equilibrium quantities for the BFM, and that the polymer's memory
effects in the BFM do not rule out our theory of translocation that
originally yielded $\tau_d\sim N^{2+\nu}$ by relating the polymer's
anomalous dynamics to its memory kernel; yet the asymptotic scaling
behavior of $\tau_d$ with $N$ for the BFM is unknown at present. (It
is the severity of the finite-size effects that makes scaling
conclusions in relation to unbiased translocation for $N\le256$
particularly meaningless). The answer to this conundrum lies in the
dynamical peculiarity of the BFM, in particular how the BFM behaves
dynamically (i.e., when $v(t)=\dot{s}(t)=0$) to the introduction of an
obstacle, which, in the case at hand is the membrane with a narrow
pore. To this end, we sample many polymer states (drawn from the
equilibrium distribution), in which the polymer is tethered halfway
through the pore with a width of two lattice spacings. For each of
these states, for each monomer, we determine the individual monomeric
mobilities. Since the orientation of the membrane breaks rotational
symmetry, we separately keep track of the moves parallel and
perpendicular to the membrane. The mobility of a monomer, parallel or
perpendicular to the membrane, is defined by the acceptance
probability of the corresponding MC move.
\begin{figure}[h]
\includegraphics[angle=270,width=0.5\linewidth]{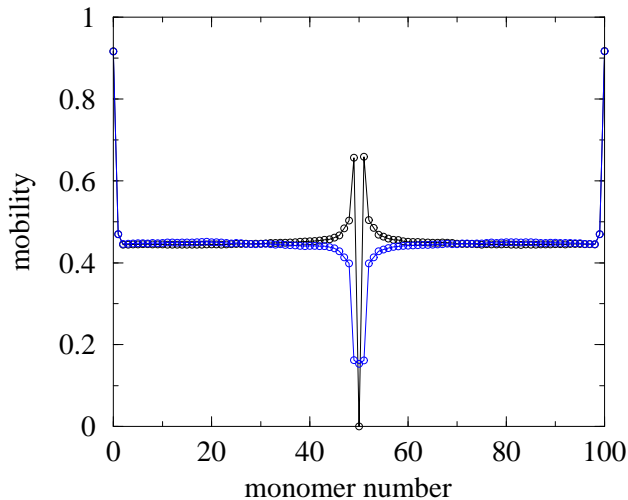}
\caption{Individual monomeric mobilities of a threaded
polymer of length $N=100$ as a function of the monomer number. The
polymer is tethered halfway through the pore, i.e., monomer number 50
is tethered in the pore. The mobility of each monomer is separately
presented in terms of the moves parallel (blue) or perpendicular
(black) to the membrane.\label{fig4}}
\end{figure}

The individual monomeric mobilities for the BFM are shown in
Fig. \ref{fig4}, for $N=100$ (with the 50th monomer tethered in the
pore). On the one hand, the monomer located in the pore shows no
mobility in the direction parallel to the membrane; this is to be
expected, since sideways mobility of this monomer is forbidden due to
the steric hindrance of the membrane. However, immediately outside the
pore, the mobility of the monomers parallel to the membrane is
strongly enhanced before it settles to a value $\approx0.44$ further
away from the pore. On the other hand, the
perpendicular-to-the-membrane mobilities of the threaded monomer and a
few (two or three) of its nearest monomers are strongly hindered. Such
anomalous behavior of the near-the-pore mobilities arises in the BFM
since the presence of the membrane results in the stretching of the
polymer around the pore, introducing an enhanced likelihood of
maximally stretched bonds, which reduce the
perpendicular-to-the-membrane mobilities of the threaded monomer and a
few of its nearest monomers. The tendency of frozen mobility
perpendicular to the membrane is peculiar for the specific types of
moves allowed in the BFM (we have also checked that this continues to
hold for longer polymers in the BFM, data not shown here). For
instance, we verified that if a collective move of two neighboring
monomers in the same direction is added to the dynamics, this tendency
of frozen mobility is removed to a large extent.  Note that in other
models, as well as in experiment, reduced mobility (``friction'') in
the pore can have various natural causes; although this can postpone
the onset of scaling, it is not expected to change scaling exponents
in the thermodynamic (long-chain) limit.  It is noteworthy to mention
here that in our lattice polymer model, in the neighborhood of the
pore, the monomeric mobilities parallel to the membrane are reduced,
while the perpendicular-to-the-membrane mobilities are marginally
enhanced (not shown here).

From a theoretical point of view, in the limit of long polymers,
either enhanced or reduced dynamics in the near vicinity of the pore
--- especially if there are only two or three monomers around the pore
that suffer from anomalous mobility problems (as in Fig. \ref{fig4})
--- should not change scaling exponents. However, such a statement is
clearly not true for the BFM, as we demonstrate below that by
enhancing the monomeric mobilities within a radius of 5 lattice sites
around the pore (roughly twice the average bond length), one can
significantly change the apparent exponent $\alpha$, towards our
predicted theoretical value $(1+\nu)/(1+2\nu)=0.7$. It is indeed
remarkable that enhancing the mobilities of typically two or three
monomers around the pore changes the apparent exponent $\alpha$
significantly for the BFM, even for fairly long polymers.

\subsection{A modified BFM with enhanced monomeric mobilities around
  the pore\label{sec5a}}

To investigate how much the anomalous mobility of the monomers around
the pore in the BFM influences the dynamics of translocation, we
perform simulations of the BFM in which all moves within a radius of 5
lattice sites around the pore are boosted by a (more or less
arbitrary)  factor 4. More precisely, all moves for all monomers
located within a distance of 5 lattice sites from the pore, either
before or after the move, are attempted four times more often than the
other moves.  Note that this does not violate detailed balance: for
every move that is oversampled, the reverse move is also oversampled.
The choice for these values in this modified BFM are both motivated by
the data of Fig. \ref{fig4}; i.e., (i) in the original BFM model,
typically only two or three monomers around the monomer that resides
in the pore at any time suffer from reduced mobility and (ii) the
factor of 4 is certainly enough to overcome even the reduced mobility
of the monomer in the pore in the perpendicular-to-the-membrane
direction, which for $N=100$ is approximately a factor 2.5 smaller
than that of the monomer far away from the pore, and increases only
slightly with polymer length. The unit of time in this modified model
is still defined by one attempted MC moves per monomer far away from
the membrane.
\begin{figure*} 
\begin{center} \begin{minipage}{0.49\linewidth}
\includegraphics[angle=270,width=\linewidth]{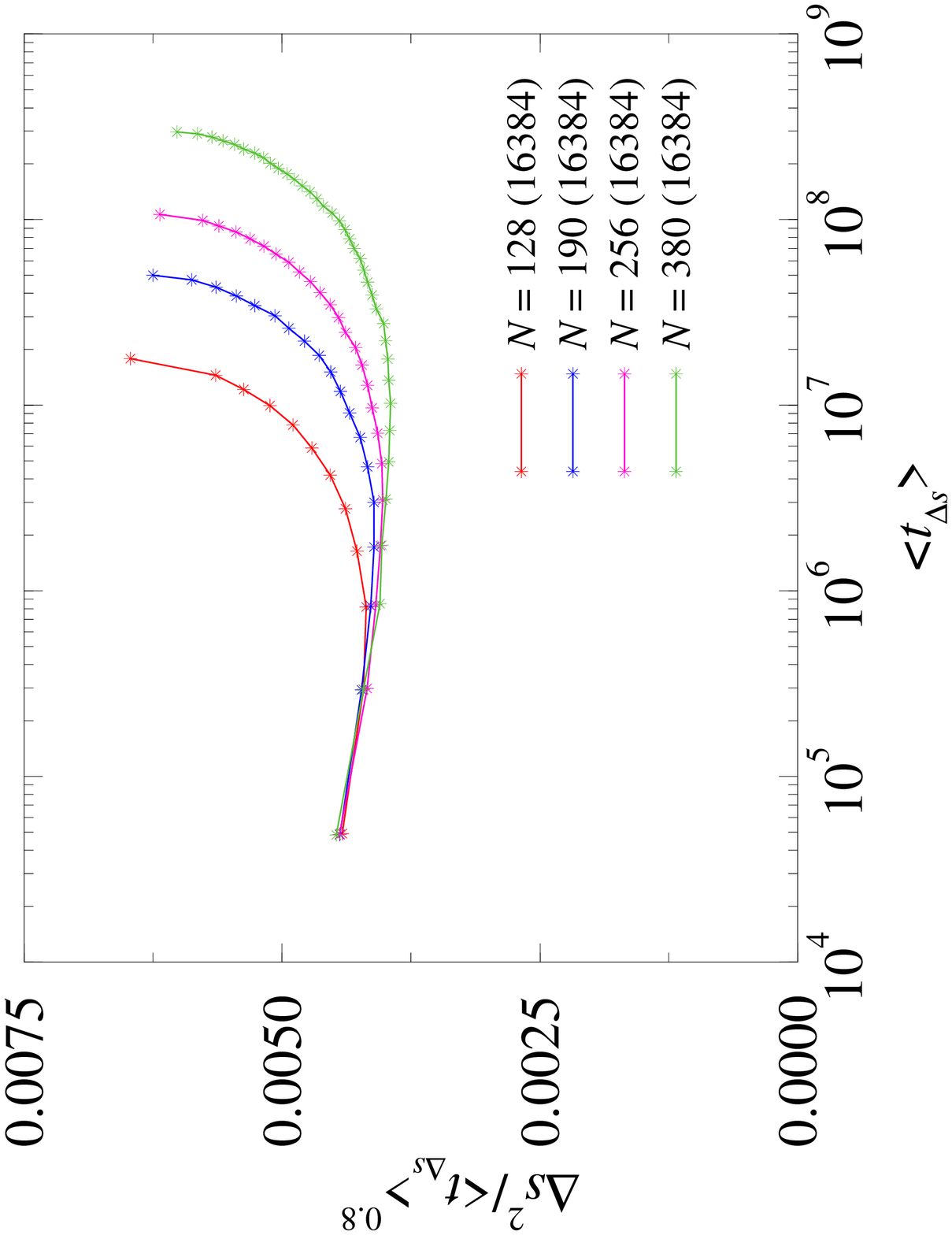}
\end{minipage} \hspace{-2mm} \begin{minipage}{0.49\linewidth}
\includegraphics[angle=270,width=\linewidth]{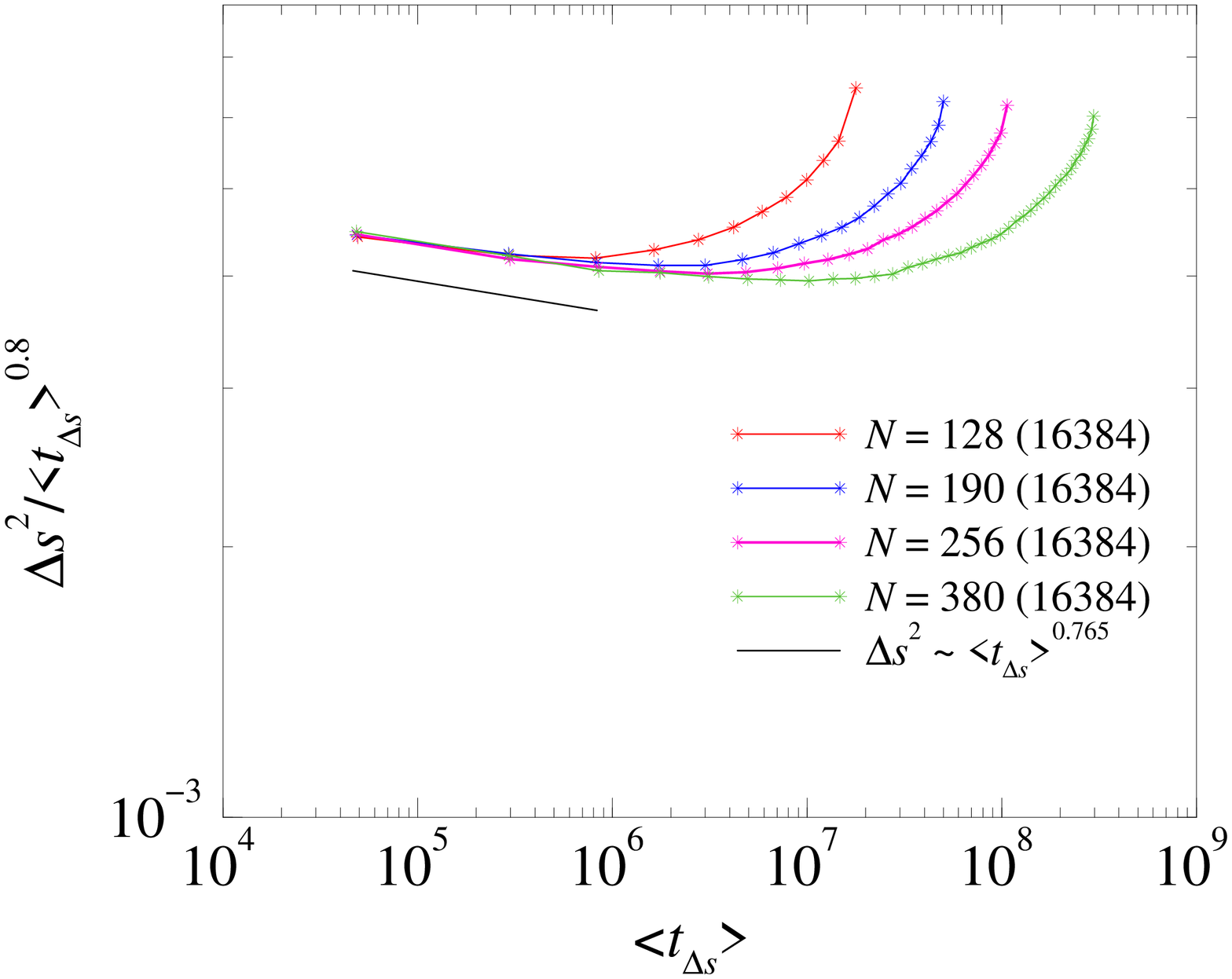} \end{minipage}
\end{center} \caption{Plots of $\Delta s^2/\langle
t_{\Delta s}\rangle^{0.8}$ as a function of $\langle t_{\Delta
s}\rangle$ with $|\Delta s|=5,10,15,\ldots,N/2$: semi-log (left
panel), log-log (right panel), analogous to Fig. \ref{fig3}. The
numbers in the parentheses in the legends denote the number of
independent realizations over which the averages have been
calculated. The lowest exponent $\alpha$ for $\langle\Delta
s^2(t)\rangle\sim t^{\alpha}$ obtained for each curve is $0.765$.
\label{fig5}}
\end{figure*}

Figure \ref{fig5} shows the anomalous dynamics of translocation ---
analogous to Fig. \ref{fig3} --- in this modified BFM: as can be seen
therein, the apparent exponent $\alpha$ has significantly decreased
(from $0.84$ in Fig. \ref{fig3} to $0.765$ in Fig. \ref{fig5}),
towards the theoretical prediction $(1+\nu)/(1+2\nu)=0.7$. The
corresponding apparent exponent $\beta\approx2.6$ up to $N=380$ (data
not shown), larger than $1+2\nu=2.5$, but smaller than $2+\nu=2.75$.

\begin{figure}[h]
\includegraphics[angle=270,width=0.6\linewidth]{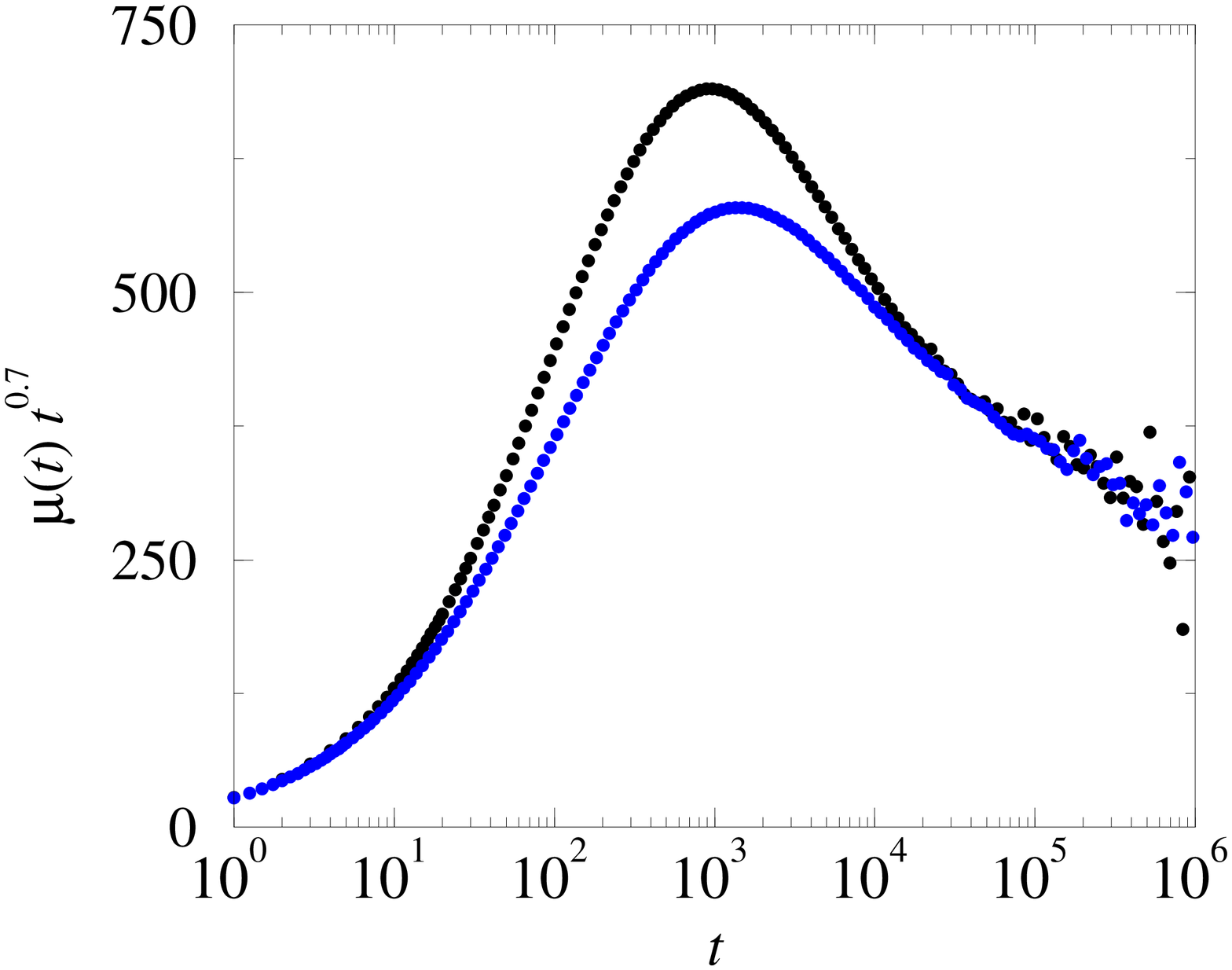}
\caption{The memory kernel $\mu(t)$ for a polymer with
length $N=1000$ for the BFM and the modified BFM. The data for the
models are averaged over $80$ and $560$ independent realizations, over
simulation times of $5\times 10^8$ and $1.25\times 10^8$ attempted
moves per monomer per realization for the BFM and modified BFM,
respectively.
\label{fignew2}}
\end{figure}

Additionally, for this modified BFM, we confirm that the memory kernel
$\mu(t-t')=\langle\phi(t)\phi(t')\rangle_{v=0}$ exhibits
$t^{-\frac{1+\nu}{1+2\nu}}$ behaviour as well. This is shown in
Fig. \ref{fignew2} for a polymer with length $N=1000$, where we also
compare the data of Fig. \ref{fignew1}. As can be seen in
Fig. \ref{fignew2}, the onset of the
$t^{-\frac{1+\nu}{1+2\nu}}=t^{-0.7}$ takes place at $t\approx1000$ for
both models.

Note that the modification we made for mobilities of the monomers in
the modified BFM is by no means what we can claim to be an exact
compensation for the around-the-pore anomalous monomeric mobilities in
the BFM, as seen in Fig. \ref{fig4}. Rather, the introduction of this
modified model should be seen as an attempt to understand whether such
anomalous mobilities can account for the deviations from the expected
values of $\beta$, namely $2+\nu=2.75$, albeit the polymer's memory
effects in the BFM confirm our theory of translocation that originally
yielded $\tau_d\sim N^{2+\nu}$ by relating the polymer's anomalous
dynamics to its memory kernel. The remarkable fact that for the BFM,
enhancing the mobilities of typically two monomers around the pore
changes the apparent exponent $\alpha$, even for fairly long polymers,
combined with the fact that the memory kernel equals
$\mu(t-t')=\langle\phi(t)\phi(t')\rangle_{v=0}\sim(t-t')^{-\frac{1+\nu}{1+2\nu}}$
at long times, leads us to conclude that the anomalous mobilities of
the monomers around the pore --- peculiarities of the BFM --- are
indeed responsible for the deviations from our expected value of
$\beta$, i.e., $2+\nu$.  We thus conclude that the BFM is {\it not\/} a
convenient model for cases where the polymer is constrained to pass
through a narrow pore.

\section{Conclusion and outlook\label{sec6}} 

In conclusion, in this paper we study unbiased polymer translocation
in two-dimensions, with the Bond Fluctuation Model (BFM), in the
absence of external forces on the polymer (i.e., unbiased
translocation) and hydrodynamical interactions (i.e., the polymer is a
Rouse polymer). While it has long been established that the
pore-blockade time $\tau_d$, the characteristic time the polymer
spends in the pore, asymptotically scales with the polymer length as
$N^\beta$ for some $\beta$, earlier studies of unbiased polymer
translocation, using the BFM, concluded that $\beta=1+2\nu$, whereas a
variety of other models produce results consistent with
$\beta=N^{2+\nu}$, originally predicted by us. Here $\nu$ is the Flory
exponent of the polymer; $\nu=0.75$ in 2D. We find that for the BFM
the quantity $f(N)=\tau_d/N^{1+2\nu}$ is a monotonically decreasing
quantity with increasing $N$, such that the rate of decrease for
$f(N)$ increases with increasing $N$. Having noted that with the BFM,
the conclusion that $\beta=1+2\nu$ has been based on simulation data
for $N\le256$, we further show in this paper that (i) the BFM suffers
from strong finite-size effects for $N\le256$, and that (ii) $f(N)$
decreases steeply for $N>256$, the conclusion that $\beta=1+2\nu$ ---
in the usual sense of critical phenomena for polymers in the limit
$N\rightarrow\infty$ --- is meaningless. We trace the peculiarities of
the BFM to the anomalously low mobility of two or three monomers in
the near vicinity of the pore, in the direction perpendicular to the
membrane.  We find that if the mobility of these monomers is enhanced,
the exponent for the pore-blockade time increases towards $2+\nu$. We
conclude that, although the BFM is a fine model for polymer dynamics
in general, it is not in situations where the polymer is constrained
to pass through a narrow pore. Our analysis also implies that for
those polymer models that assert $\beta=1+2\nu$, one needs to
thoroughly investigate their finite-size effects and dynamical
peculiarities, if such an assertion is to be proved meaningful.

A related issue regarding the use of the BFM for translocating
polymers does still remain, and that is the case of field-driven
translocation in two dimensions. For this situation, translocation is
driven by a potential difference across the pore. (All results quoted
below are for Rouse polymers.) First of all, using the same memory
effects as in unbiased translocation, in Ref. \cite{panja4} we argued
that in 3D, the pore-blockade time exponent is
$(1+2\nu)/(1+\nu)\approx1.37$, and corroborated this result with
extensive simulations; this result has now been verified with
completely different polymer models \cite{luo08,fyta}. Further, in
Ref. \cite{panja2}, we argued that for field-driven translocation in
2D, because of energy conservation, the pore-blockade time exponent
has a lower bound $2\nu=1.5$: for any field strength, a polymer length
exists above which the Rouse friction prevents the transport of
monomers from keeping up with the speed of translocation dictated by
the memory effects. The exponent $(1+2\nu)/(1+\nu)$ is therefore not
observed for field-driven translocation in 2D. We also showed,
numerically, using our lattice polymer model that in 2D the
pore-blockade time exponent indeed turns out to be the same as its
lower bound $2\nu$ \cite{panja2}, which is in agreement with those
obtained by the use of the BFM by two separate research groups
\cite{kantorf,luijten}, who used $N$ up to $600$ and $256$
respectively. However, a third group which also used the BFM claimed
the pore-blockade time exponent for field-driven translocation in 2D
to be consistent with $2\nu$ for $N$ up to $300$, and $1+\nu$ for
$N>300$ \cite{luo10,luoold}, which is clearly at odds with the results
of Refs. \cite{panja2,kantorf,luijten}. If indeed the BFM finds the
pore-blockade time exponent to be $1+\nu$, then the anomalous
mobilities of the monomers around the pore, together with the
finite-size effects, may explain why the exponent $2\nu=1.5$ is not
observed in the BFM for field-driven translocation in two dimensions.

\noindent{\bf Acknowledgments:} D. P. acknowledges ample computer
time on the Dutch national supercomputer facility SARA.


\begin{thebibliography}{08}
\bibitem{comment} Since chain tension is not a directly measurable
observable in the BFM, we considered the perpendicular-to-the-membrane
distance $Z_m$ of the center-of-mass of the first $m$ monomers as a
proxy for the chain tension [see for example Ref. \cite{panjabj},
where we used $m=4$]. To verify that the local extension is indeed
proportional to the chain tension at the pore, we performed separate
simulations (not reported), in which we pulled on the free end of a
tethered chain with a known (small) force $f$, and recorded $\langle
Z_m\rangle$. It turned out that for $m=1$ or 2, $\langle Z_m\rangle$
shows slightly non-linear behavior as a function of $f$ for small forces,
while for $m=4$ or higher, $\langle Z_m\rangle$ is linear in $f$ for
small and medium-strength forces.
\end{thebibliography}
\end{document}